\documentclass[a4paper]{article}
\usepackage{graphicx}
\usepackage{amsmath}

\input{tcilatex}

\begin{document}

\title{Solvent evaporation of spin cast films:\\
''crust'' effects}
\author{P.\ G.\ de Gennes \\
%EndAName
Coll\`{e}ge de France, 11 place M.\ Berthelot\\
75231 Paris Cedex 05, France\\
E.mail: pgg@espci.fr}
\maketitle

\begin{abstract}
When a glassy polymer film is formed by evaporation, the region near the
free surface is polymer rich and becomes glassy first, as noticed long ago
by Scriven et al. We discuss the thickness of this ''crust'' and the time
interval where it is present -before freezing of the whole film.\ We argue
that the crust is under mechanical tension, and should form some cracks.\
This may be the source of the roughness observed on the final, dry films,
when the solvent vapor pressure is high (and leads to thin crusts).

\bigskip

\textit{PACS numbers: 66.10-x; 68.60-p; 83.80.Rs; 82.70-y.}

\bigskip

\textit{Shortened version of the title}:

\bigskip

\ \ \ \ \ \ \ \ \ \ \ \ \ \ \ \ \ \ \ \ \ \ \ \ \ \ SPIN\ CAST\ FILMS
\end{abstract}

\section{Introduction}

Spin cast polymer films are used in many industrial sectors (electronics,
packaging, ...).\ But the birth of the films is complex: during the (rapid)
solvent evaporation, many things happen. In particular, one can think of:

\qquad a) thermal (Rayleigh Benard) instabilities (since the free surface is
cooled down).

\qquad b) convective instabilities due to concentration effects (when the
surface tension of the polymer $\gamma _{p}$ is higher than the surface
tension of the solvent $\gamma _{s}$): a solvent rich plume lowers the
surface tension, and this enhances the plume.

We recently argued \cite{pgg} that (when $\gamma _{p}>\gamma _{s}$) process
(b) should dominate over process (a).

In the present note, we are concerned with another phenomenon.\ Strawhecker
et al \cite{strawhecker} found that the surface roughness of the final films
is anomalously high ($\sim 50nm$) when the pure solvent has a high vapor
pressure.

This cannot be explained by process (b) above: a number of polymer solvent
pairs with $\gamma _{p}>\gamma _{s},$ give a smooth surface (e.g. PS/toluene
on PVME/water).\ Also, some systems with $\gamma _{p}<\gamma _{s}$, give a
rough surface (e.g. PS/acetone or PVME/methanol).

This led us to another line of thought, based on the glassy nature of the
final state: a polymer rich ''crust'' builds up near the free surface: when
it dries out, it is under tension and it should rupture -creating a rough
surface.

In section 2, we discuss the concentration profiles in the film, and the
formation of the crust. This has been analysed many years ago in precise
numerical \ calculations by Bornside, Macosko and Scriven \cite{bornside}.\
Here we set up a much cruder, but more transparent, model. In section 3, we
produce a crude estimate of the mechanical tensions, and discuss the
possible forms of rupture. All our analysis is qualitative: any improvement
on this would require a deep (non existent) knowledge of the glass
transition induced by solvent depletion.

\section{Crust formation}

\subsection{Transport in air}

The aspect of the concentration profiles at one, given instant $t$ during
evaporation is shown on fig.\ref{fig1}.\ The solvent volume fraction $\psi $
has a high value $\psi _{d}(t)$ at the bottom plate, and a low value $\psi
_{u}(t)$ at the free surface. Immediately above this, we have $\psi =\psi
_{g}$. This value corresponds to a solvent partial pressure $p_{g}$ in the
neighboring gas. Inside the gas, we assume a diffusion layer of fixed
thickness $\ell $: this is a crude approximation to the actual boundary
layers which are associated with air motions in the laboratory.

\FRAME{fhFU}{2.7017in}{2.0435in}{0pt}{\Qcb{solvent volume fractions near the
first freezing time ($t=t^{\ast }$) (qualitative picture)}}{\Qlb{fig1}}{%
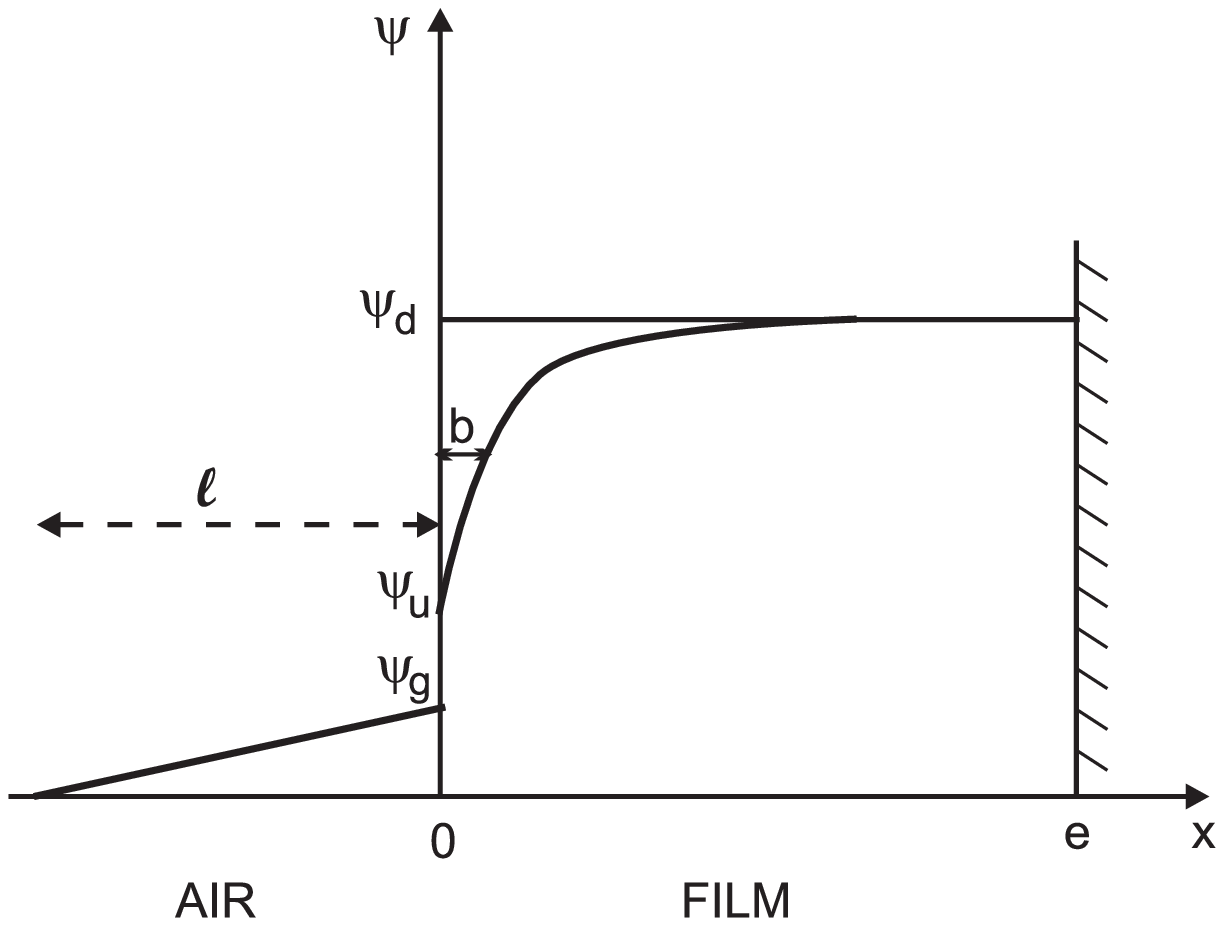}{\special{language "Scientific Word";type
"GRAPHIC";maintain-aspect-ratio TRUE;display "PICT";valid_file "F";width
2.7017in;height 2.0435in;depth 0pt;original-width 4.862in;original-height
3.6642in;cropleft "0";croptop "1";cropright "1";cropbottom "0";filename
'fig12.eps';file-properties "XNPEU";}}

The outward solvent current (in air) is related to a diffusion coefficient $%
D_{air}:$

\begin{equation}
J=D_{air}\frac{p_{\text{v}}}{kT}\frac{1}{\ell }  \label{eq1}
\end{equation}

Here, $p_{\text{v}}/kT\equiv \phi _{g}/a^{3}$ is the number density of
solvent just above the interface, \ and $J$ is a number of molecules per
unit area and unit time. The diffusion constant $D_{air}$ is of order v$%
_{th}\lambda $, where v$_{th}=(kT/m)^{1/2}$ is the thermal velocity for
solvent molecules of mass $m$, and $\lambda $ the mean free path in air
(inversely\ proportional to the atmospheric pressure $p_{a}$).\ This gives
ultimately:

\begin{equation}
D_{air}\sim \frac{\text{v}_{th}}{a^{2}}\frac{kT}{p_{a}}  \label{eq2}
\end{equation}

where $a$ is the size of a solvent molecule.

We are thus led to an evaporation current:

\begin{equation}
J=\frac{\text{v}_{th}}{a^{2}\ell }\frac{p_{g}}{p_{a}}  \label{eq3}
\end{equation}

\subsection{Local equilibrium at the free surface}

We assume, for simplicity, that the volume fractions just below ($\psi _{u}$%
) and just above ($\psi _{g}$) the interface are related by Henry's law,
with a constant coefficient:

\begin{equation}
a^{3}\psi _{g}\equiv \frac{p_{g}}{kT}=\psi _{u}\frac{p_{\text{v}}(T)}{kT}
\label{eq4}
\end{equation}

where $p_{\text{v}}$ is the vapor pressure of pure solvent.\ This assumption
ignores many delicacies in the sorption\ desorption curves \cite{leibler}, 
\cite{saby}, but it should be sufficient for our purposes.\ Eqs (\ref{eq3}, 
\ref{eq4}) then give us:

\begin{equation}
\frac{J}{a^{3}}=\frac{\text{v}_{th}a}{\ell }\frac{p_{\text{v}}(T)}{p_{a}}%
\psi _{u}  \label{eq5}
\end{equation}

(where $a^{3}$ is the volume per solvent molecule in the liquid solvent).

\subsection{Steady state currents in the crust region}

Inside the crust, we assume that a steady state is achieved, with the same
current $J$. If we call $D(\phi )$ the diffusion coefficient of the solvent
in the mixture, we may write:

\begin{equation}
J=D(\psi )\frac{d\psi }{dx}  \label{eq6}
\end{equation}

The coefficient $D(\psi )$ varies with $\psi $ for two reasons:

\qquad a) the mesh size $\xi $ of the polymer solution increases with $\psi $
$\cite{pgg2}.$

\qquad b) at low $\psi $ the system is glassy and $D(\psi )$ becomes very
small.

Effect (a) is minor compared with effect (b), and we shall omit it in the
following. To get a practical feeling about $D(\psi )$ in the small $\psi $
region, it is helpful to use a free volume picture for the glass transition 
\cite{grest}.\ The free volume parameter v has one component v$_{1}$ present
in pure polymer and another component proportional to the volume fraction of
solvent:

\begin{equation}
\frac{\text{v}(T_{1}\phi )}{a^{3}}=\frac{\text{v}_{1}(T)}{a^{3}}+k\psi
\label{eq7}
\end{equation}

(where $k$ is a coefficient of order unity).

We shall focus our attention on the region where solvent effects dominate
over temperature effects ($\psi >$v$_{1}/a^{3}).\;$This then gives:

\begin{equation}
D(\psi )=D_{1}\exp \left( \frac{q}{\psi }\right)  \label{eq8}
\end{equation}

where $q$ is another coefficient of order unity, and $D_{1}$ is the
diffusion constant in a very fluid mixture ($\psi \sim 1$).\ Where we know
an explicit form of $D(\psi )$ such as eq. (\ref{eq8}), we can find the
steady state profile in the crust region by integrating eq. (\ref{eq7})
over, as thickness $x$ near the free surface:

\begin{equation}
x=J\int_{\psi _{u}}^{\psi }D(\psi ^{\prime })d\psi ^{\prime }  \label{eq9}
\end{equation}

Eq. (\ref{eq9}) holds only in the region of small $D$, which acts as a
barrier, when the assumption $\partial J/\partial x=0$ is valid. Outside of
the barrier, diffusion is fast and the profile $\psi (x)$ is nearly flat.

Note that the presence of a crust, as it is understood here, is independent
of the presence (or absence) of an adsorbed polymer layer near the free
surface.\ An adsorbed layer will be present if $\gamma _{p}<\gamma _{s}$,
and if the adsorption time is shorter \cite{pgg2} than the overall time for
spin casting.\ But this layer is expected to be very thin; the number $%
\Gamma $ of adsorbed monomers per unit area should be $\Gamma \sim a^{-2}$:
a thin layer like this cannot be really glassy, and does not contribute
significantly to the crust.

\subsection{The crust thickness}

We can now define the thickness of the crust $b$ via the initial slope at $%
x=0$ (the free surface):

\begin{equation}
\frac{1}{b}=\left. \frac{1}{\psi _{u}}\frac{d\psi }{dx}\right| _{x=0}
\label{eq10}
\end{equation}

Using eqs (\ref{eq5}, \ref{eq6}), this gives:

\begin{equation}
b\cong \ell \frac{p_{a}}{p_{\text{v}}}\frac{D(\psi _{u})}{\text{v}_{th}a}
\label{eq11}
\end{equation}

The solvent fracture $\psi _{u}(t)$ decreases with time: at a certain
instant $t^{\ast }$, it reaches a critical value $\phi ^{\ast }$ below which
the polymer solvent system is glassy. The central parameter is the crust
thickness $b^{\ast }$ at this moment:

\begin{equation}
b^{\ast }=\ell \frac{p_{a}}{p_{\text{v}}}\frac{D(\psi ^{\ast })}{\text{v}%
_{th}a}  \label{eq12}
\end{equation}

For instance, let us use the general form \ref{eq8} for $D(\psi )$, with $%
q=1 $ and $\psi ^{\ast }=0.2.\;$Taking $\ell =1mm,$ $p_{a}/p_{\text{v}}=10,$
and $Do/a$v$_{th}=10^{-3}$, we arrive at $b^{\ast }=70$ nanometers.\ Thus
the crust is indeed thin for practical conditions.

Another important feature of eq. (\ref{eq12}), is the dependence on the
vapor pressure $p_{\text{v}}(T)$: \textit{large vapor pressures lead to thin
crusts}.\ This, in turn, implies that the crust will be more fragile, as
discussed in section 3.

\subsection{Lifetime of the crust}

The birth of the crust occurs at a certain time $t^{\ast }$ (when $\psi
_{u}=\psi ^{\ast }$).\ At a later time $t^{\ast \ast }$, the whole film
becomes glassy (when $\psi _{d}=\psi ^{\ast }$).\ We shall now assume that $%
t^{\ast \ast }-t^{\ast }$ is (like $t^{\ast }$), proportional to the overall
evaporation time $\tau _{e\text{v}}$.\ A simplified discussion of $\tau _{e%
\text{v}}$ is given below.

We concentrate on regimes where $b$ is smaller than the overall thickness $%
e(t)$.\ Then, the total amount of solvent in the film is $Q\cong e\psi _{d}$
(per unit area of film).

The rate of change of $Q$ is given by:

\begin{equation}
\frac{dQ}{dt}=-\frac{J}{a^{3}}+\phi _{u}\frac{de}{dt}  \label{eq13}
\end{equation}

where the first term describes evaporation, while the second term is related
to the presence of a moving boundary.

The conservation of polymer imposes:

\begin{equation}
e-Q\equiv e(1-\psi _{d})=e_{f}  \label{eq14}
\end{equation}

where $e_{f}$ is the final thickness of the dry film.

Eqs (\ref{eq13}, \ref{eq14}) must still be supplemented by one relation
relating $\psi _{d}$ to $\psi _{u}$.\ We shall obtain this in a very crude
fashion, by considering the cross over point between crust and ''inside''
the film. We assume that at $\psi =c\psi _{u}$ (where $c$ is a numerical
coefficient of order 2), the diffusion coefficient $D(\psi )$ has become
fast: then the profile is flat, and this implies $\psi =\psi _{d}$.\ Thus,
we are led to the ansatz:

\begin{equation}
\psi _{d}=c\psi _{u}  \label{eq15}
\end{equation}

Then the system (\ref{eq13}-\ref{eq15}) can be reduced to:

\begin{equation}
ced\psi _{u}+de(c-1)\psi _{u}=-\frac{J}{a^{3}}dt=-A\psi _{u}dt  \label{eq16}
\end{equation}

with:

\begin{equation}
A=\frac{\text{v}_{th}a}{\ell }\frac{p_{a}}{p_{\text{v}}}  \label{eq17}
\end{equation}

Eq. (\ref{eq16}) together with eq. (\ref{eq14}) can be integrated in
detail.\ But, for our purposes, it is enough to note that it involves a
single time constant $\tau :$

\begin{equation}
\tau _{e\text{v}}=c^{-1}\frac{e_{f}}{A}  \label{eq18}
\end{equation}

In the form (\ref{eq18}), $\tau $ is exactly the relaxation time in the
final stage, where $e\rightarrow e_{f}$ and $de/dt$ can be neglected.

Thus, we are led to postulate that the duration of the crust regime follows
the scaling rule:

\begin{equation}
t^{\ast \ast }-t^{\ast }\sim \tau _{e\text{v}}\sim \frac{e_{f}}{A}\sim \frac{%
e_{f}\ell }{\text{v}_{th}a}\frac{p_{a}}{p_{\text{v}}}  \label{eq19}
\end{equation}

\section{Rupture of the crust and resulting effects}

We now focus our attention on the interval $t^{\ast }<t<t^{\ast \ast }$,
where the crust is present over a sheet of fluid solution. During this
interval, the volume fraction $\psi _{u}(t)$ at the free surface decreases
from $\psi ^{\ast }$ down to a finite fraction of $\psi ^{\ast }$ -say $\psi
^{\ast }/c$: this would ensure that the volume fraction at the bottom plate $%
\psi _{d}$ reaches the threshold $\psi _{d}(t^{\ast \ast }):\psi ^{\ast }$
at the end of the interval.

Following the ideas of Leibler and Sekimoto \cite{leibler}, we believe that
a network is formed in the crust as soon as $\psi _{u}(t)$ reaches $\psi
^{\ast }.$ At later times, this network is deswollen (as explained on fig. 
\ref{fig2}), because $\psi $ decreases down to $\psi ^{\ast }/c$. The volume
of the gel decreases, but its horizontal dimensions have to remain the same.
Thus, there is a tensile stress in the crust.

\FRAME{fhFU}{3.1868in}{3.243in}{0pt}{\Qcb{the ''crust'': \ \ a) overall view
\ \ b) enlarged view of one mesh unit at the first freezing time ($t=t^{\ast
})$ \ \ c) the same unit, at later times ($t\sim t^{\ast \ast }$), is under
horizontal tension.}}{\Qlb{fig2}}{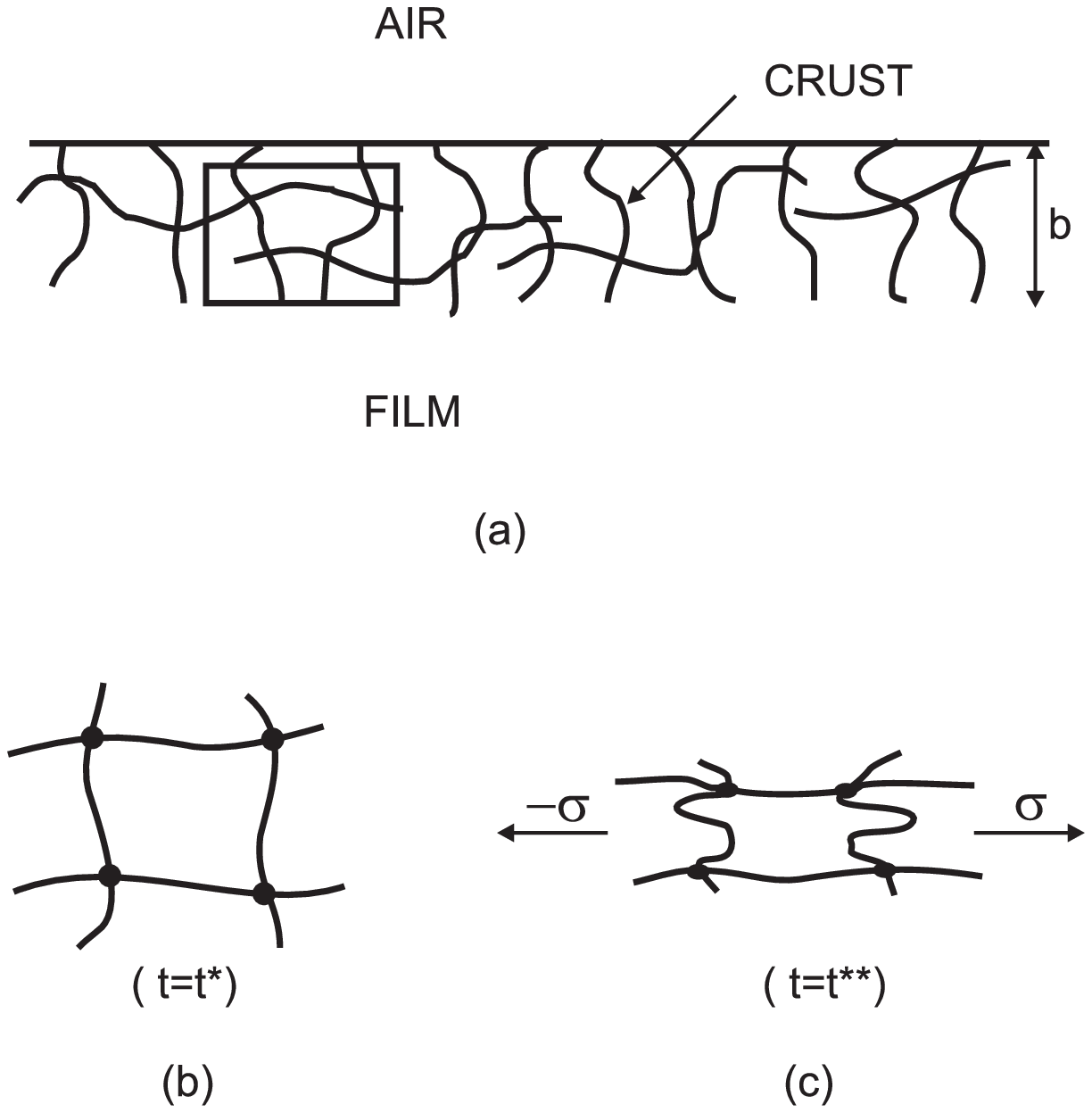}{\special{language "Scientific
Word";type "GRAPHIC";maintain-aspect-ratio TRUE;display "PICT";valid_file
"F";width 3.1868in;height 3.243in;depth 0pt;original-width
4.8594in;original-height 4.9467in;cropleft "0";croptop "1";cropright
"1";cropbottom "0";filename 'fig24.eps';file-properties "XNPEU";}}

The overall contraction ratio is of order:

\begin{equation}
\theta =-\psi _{u}(t^{\ast \ast })+\psi _{u}(t^{\ast })\cong \psi ^{\ast
}\left( 1-\frac{1}{c}\right) \cong \psi ^{\ast }  \label{eq20}
\end{equation}

and the tensile shears stress $\sigma $ is predicted to be:

\begin{equation}
\sigma \cong \Pi (\psi ^{\ast \ast })\theta =\Pi (\psi ^{\ast \ast })\psi
^{\ast }  \label{eq21}
\end{equation}

where we have replaced the elastic\ modulus of the gel by the osmotic
pressure of the polymer $\Pi $, which scales in the same way.\ Because at $%
\psi =\psi ^{\ast \ast }$ we are dealing with rather concentrated solutions,
we may write simply \cite{pgg2}:

\begin{equation}
\Pi (\psi ^{\ast \ast })\cong \frac{kT}{a^{3}}  \label{eq22}
\end{equation}

The ratio of stress to elastic modulus is thus expected to rise up to a
value:

\begin{equation}
\frac{\sigma }{\Pi (\psi ^{\ast \ast })}\sim \psi ^{\ast }  \label{eq23}
\end{equation}

With the usual values of $\psi ^{\ast }$ (0,2), this should be sufficient to
induce fracture in the crust.\ Fracture is indeed favored because \ \ a) the
crust is thin \ \ b) it must have the usual syneresis, leading to rather
open regions which can play the role of a nucleation center.

Thus, we are led to suggest that the outer surface of the film exhibits a
network of fracture lines at times $t<t^{\ast \ast }$, very much like earth
at the bottom of a drying pond. The evaporation flows should then converge
towards these fracture lines, and a rough interface should appear. We are
not able to predict the density of cracks, nor the amplitude of the
resultant roughness. But it is clear that this process should be important,
mainly when the crust is thin.\ By eq. (\ref{eq12}) this requires a high
vapor pressure $p_{\text{v}}$ for the solvent -in agreement with the results
of ref. \cite{strawhecker}.

How could we test these ideas?

\qquad a) apart from acting on $p_{\text{v}}$, we could also possibly act on
the boundary layer conditions (described by $\ell $ in eq. \ref{eq12}), or
decrease the air pressure.

\qquad b) some direct observation may be possible. Crusts have been observed
in deswelling experiments by Tanaka and coworkers \cite{tanaka}, using
optical microscopes. They did see crust rupture after some deswelling. In
the present case, it may be better to probe the surface (after complete
drying) with an AFM\ tip.

Cracks are the most plausible channel for relaxation of the tension.\ But
the following evaporation flow may lead to other forms -e.g. ''volcanic''
landscapes.

To summarize: all our discussion is extremely qualitative, and it may be
insufficient at some points.\ However, we believe that \ \ a) the compact
formula (\ref{eq12}) for the crust thickness may be useful \ \ b) crust
rupture is indeed a plausible explanation for the surface roughness of the
films.

\bigskip

\textit{Acknowledgments:} I benefited from very stimulating exchanges with
S.\ K.\ Kumar, G.\ Reiter and F.\ Brochard-Wyart.

\end{document}